\documentclass[%
reprint,
superscriptaddress,
amsmath,amssymb,
aip
]{revtex4-2}

\usepackage{graphicx}
\usepackage[dvipsnames]{xcolor}
\usepackage{dcolumn}
\usepackage{bm}
\usepackage[utf8]{inputenc}
\usepackage{siunitx}
\usepackage[T1]{fontenc}
\usepackage{amsmath}

\draft 

\begin{document}

\title[Thompson et al.]{Voltage-tunable Josephson Junctions on Germanium Quantum Wells with \textit{in-situ} Aluminum Contacts}

\author{J. P. Thompson}
\email[]{jpthomp@umd.edu}
\altaffiliation{These authors contributed equally to this work.}
\affiliation{Laboratory for Physical Sciences, 8050 Greenmead Drive, College Park, MD 20740, USA}
\affiliation{Department of Physics, University of Maryland, College Park, MD 20742, USA}

\author{J. T. Dong}
\altaffiliation{These authors contributed equally to this work.}
\affiliation{Laboratory for Physical Sciences, 8050 Greenmead Drive, College Park, MD 20740, USA}
\affiliation{Department of Physics, University of Maryland, College Park, MD 20742, USA}

\author{B. Langa Jr.}
\affiliation{Laboratory for Physical Sciences, 8050 Greenmead Drive, College Park, MD 20740, USA}
\affiliation{Department of Physics, University of Maryland, College Park, MD 20742, USA}

\author{C. K. Gaspe}
\affiliation{Laboratory for Physical Sciences, 8050 Greenmead Drive, College Park, MD 20740, USA}

\author{R. Card}
\affiliation{Laboratory for Physical Sciences, 8050 Greenmead Drive, College Park, MD 20740, USA}
\affiliation{Department of Materials Science \& Engineering, University of Maryland, College Park, MD 20742, USA}

\author{B. Bailey}
\affiliation{University of Arkansas, Department of Physics, 226 Physics Building
825 West Dickson Street, Fayetteville, Arkansas 72701, USA}

\author{S. Davari}
\affiliation{University of Arkansas, Department of Physics, 226 Physics Building
825 West Dickson Street, Fayetteville, Arkansas 72701, USA}

\author{B. E. Matthews}
\affiliation{Energy and Environment Directorate, Pacific Northwest National Laboratory, Richland, Washington 99352, USA}

\author{M. J. Olszta}
\affiliation{Energy and Environment Directorate, Pacific Northwest National Laboratory, Richland, Washington 99352, USA, USA}

\author{S. Hoffman}
\affiliation{Laboratory for Physical Sciences, 8050 Greenmead Drive, College Park, MD 20740, USA}

\author{T. M. Hazard}
\affiliation{Lincoln Laboratory, Massachusetts Institute of Technology, Lexington, MA 02421, USA}

\author{K. Serniak}
\affiliation{Lincoln Laboratory, Massachusetts Institute of Technology, Lexington, MA 02421, USA}

\author{H.O.H. Churchill}
\affiliation{University of Arkansas, Department of Physics, 226 Physics Building
825 West Dickson Street, Fayetteville, Arkansas 72701, USA}

\author{K. Sardashti}
\affiliation{Laboratory for Physical Sciences, 8050 Greenmead Drive, College Park, MD 20740, USA}
\affiliation{Department of Physics, University of Maryland, College Park, MD 20742, USA}

\author{C.~J.~K.~Richardson}
\affiliation{Laboratory for Physical Sciences, 8050 Greenmead Drive, College Park, MD 20740, USA}
\affiliation{Department of Materials Science \& Engineering, University of Maryland, College Park, MD 20742, USA}

\date{\today}

\begin{abstract}
Voltage-tunable Josephson junctions (VT-JJs) are an emerging element in superconducting quantum electronics with potential to expand the functionality of conventional designs. While VT-JJs are largely compatible with wafer-scale semiconductor processing, their integration into quantum circuits remains a challenge due to unmitigated semiconductor microwave loss. Here, a deep mesa etch process, wherein the epitaxial material is removed except the VT-JJ device, will facilitate the integration of VT-JJs with low-microwave-loss circuit elements by allowing these circuit elements to be placed directly on a low-loss substrate. A Germanium quantum well is grown by Molecular Beam Epitaxy (MBE) on a float zone silicon substrate with \textit{in-situ} deposited aluminum contacts. This combination allows the formation of an oxide-free superconductor-semiconductor interface. The deep mesa etch process is optimized to produce a sidewall taper sufficient for continuous metal deposition from the substrate to the top of the mesa for electrostatic gate electrodes and interconnects. The fabricated Josephson junctions demonstrate gate-tunable supercurrents with a maximum critical current over 100~\unit{nA} and critical-current normal-resistance product of 8.63~\unit{\uV}. These results demonstrate a pathway toward improved integration of voltage-tunable superconducting circuit elements with quantum electronic building blocks such as couplers and qubits.

\end{abstract}


\maketitle 

In recent years, voltage-tunable Josephson Junctions (VT-JJs) have emerged as promising building blocks for scalable quantum processors. A key functionality that VT-JJs offer over traditional Al/AlO\textsubscript{x}/Al JJs is electrostatic tuning of supercurrents \cite{kleinsasser_superconducting_1989, akazaki_josephson_1996, aguado_perspective_2020}. Voltage tuning may eliminate the need for flux lines currently used to modulate frequencies in superconducting circuits, thereby reducing a potential source of flux noise and reducing line counts and improve thermal performance \cite{paladino_1f_2014,rosenberg_3d_2017, krantz_quantum_2019}. Voltage tuning is made possible by replacing the insulating tunnel barrier with a semiconductor.  An applied electric field can control the supercurrent flowing across the junction, allowing the Josephson energy to be tuned. This key functionality of VT-JJs motivated the demonstrations of multiple voltage-tunable superconducting quantum devices such as gatemon qubits \cite{larsen_semiconductor-nanowire-based_2015, de_lange_realization_2015, casparis_superconducting_2018}, voltage-tunable couplers \cite{schmidt_ballistic_2018, sardashti_voltage-tunable_2020, hazard_superconducting-semiconducting_2023}, voltage-tunable fluxonium qubits \cite{strickland_gatemonium_2025}, and Andreev qubits\cite{hays_coherent_2021, pita-vidal_direct_2023,cheung_photon-mediated_2024}. To date, VT-JJs have been successfully realized on multiple hybrid superconductor-semiconductor (S-Sm) materials platforms from nanowires to 2D flakes and planar quantum wells (QWs) \cite{de_lange_realization_2015,zhuo_hole-type_2023, zheng_coherent_2024,wang_coherent_2019, casparis_superconducting_2018, strickland_characterizing_2024, sagi_gate_2024,kiyooka_gatemon_2025}. However, full integration of VT-JJs into highly coherent quantum circuits requires the material platform to be synthesized at large scale and be fabricated entirely from low-loss microwave materials. Among these material systems, planar QWs on float-zone silicon are well suited for wafer-scale manufacturing processes due to their compatibility with fabrication techniques from the semiconductor industry.

An effective VT-JJ also requires ballistic transport across the junction and high interface transparency of the supercurrent across the superconductor-semiconductor interface. Obtaining high interface transparency between the superconductor and QW is accomplished by proximitized superconductivity in the QW that is directly below the superconductor. High interface transparency can be identified by the JJ expressing a hard superconducting gap, and it has been a materials challenge \cite{krogstrup_epitaxy_2015, chang_hard_2015} . One approach applied to planar QWs is to anneal the device in order to promote diffusion of the superconducting contacts into the SiGe \cite{ridderbos_hard_2020, hendrickx_ballistic_2019} or form a superconducting germanosilicide \cite{tosato_hard_2023}. However, annealing can be difficult to precisely control because it can readily introduce nanometer-scale non-uniformity in the junction gap length from lateral diffusion of the contact metal into the tunnel region. 

In order to improve VT-JJs for superconducting quantum circuits, scalable integration with low-loss substrates is required. 
 Implementing VT-JJs with planar Ge QWs is routinely achieved by growing the SiGe structure on undoped float-zone Si substrates with microwave loss tangents $\leq 10^{-6}$. Demonstrated here is a voltage-tunable JJs fabricated from molecular beam epitaxy (MBE)-grown material. Shallow Ge QWs with \textit{in-situ} deposited Al contacts achieve JJs with ballistic transport across the junction. A deep mesa etch is implemented with features on and off the mesa to demonstrate a path towards fully removing the lossy epitaxial buffer layers in regions away from the JJ in order to minimize microwave loss \cite{sandberg_investigating_2021}.

\begin{figure*}[ht]
\centering
\includegraphics[width=\textwidth]{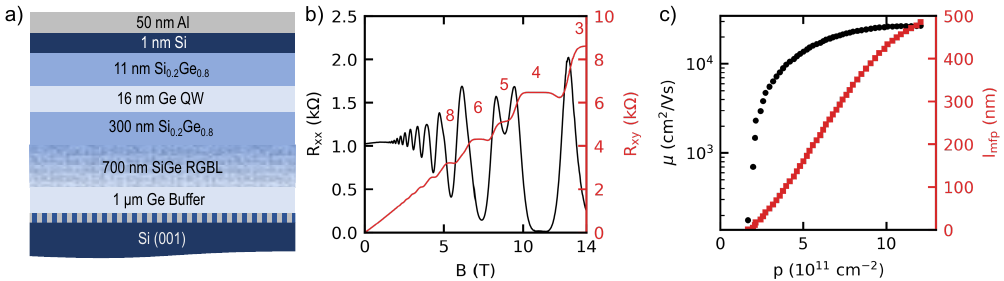}
\caption{\label{fig:1} a) Al/Ge QW heterostructure schematic. b) Magnetotransport for a Hall bar with hole density, $\rho_H = 1.2 \times 10^{12}$~cm\textsuperscript{-2}. The indexes of the plateaus in R\textsubscript{xy} are labeled. c) Low temperature mobility (left axis) and hole mean free path (right axis) at various carrier densities, with a peak mobility of 26,800~cm\textsuperscript{2}/Vs and a peak mean free path of 482 nm. Both measurements were completed at a temperature of 2 K.}
\end{figure*}

Low-loss, float-zone silicon (001) substrates were used for this study. The substrates were etched with 5 \% hydrofluoric acid prior to loading in the MBE. The passivating hydrogen as well as residual oxygen and fluorine were desorbed from the surface at 800~\unit{\degreeCelsius}. A 150-nm-thick Si homoepitaxial buffer layer was grown at 600~\unit{\degreeCelsius}. Subsequently, a 1-\unit{\um}-thick abrupt metamorphic Ge buffer was grown at 300~\unit{\degreeCelsius} and annealed at 750~\unit{\degreeCelsius} to reduce the threading dislocation density. The sample was then cooled to 300 \unit{\degreeCelsius}. Without pausing growth, a 700-\unit{\nm}-thick SiGe reverse graded buffer with the composition linearly graded from pure Ge to Si\textsubscript{0.2}Ge\textsubscript{0.8}, a 300-nm-thick Si\textsubscript{0.2}Ge\textsubscript{0.8} bottom barrier, a 16-nm-thick Ge QW, an 11-nm-thick Si\textsubscript{0.2}Ge\textsubscript{0.8} top spacer, and a 1-nm-thick Si cap were grown. After the growth of the entire SiGe heterostructure, the sample was transferred, \textit{in-situ}, to another interconnected MBE chamber and 50-nm of Al was grown at 10~\unit{\degreeCelsius} on the surface of the Si cap. A schematic layer structure is shown in Fig. \ref{fig:1}a). 

The transport properties of the Ge QWs was characterized with gated Hall bars at 2 K (Fig. \ref{fig:1}b). The observation of the integer quantum Hall effect with vanishing magnetoresistance in the dip at B = 11 \unit{T}, indicates that the samples are of high quality and devoid of parallel conduction. The mobility of the Ge QW as a function of carrier density is shown in Fig. \ref{fig:1}c), with the different carrier densities obtained by varying the gate voltage. A peak mobility of 26,800~cm\textsuperscript{2}/Vs is measured at carrier density, $\rho_H = 1.2 \times 10^{12}$~cm\textsuperscript{-2}. These mobilities are comparable to shallow Ge QWs grown by chemical vapor deposition (CVD)\cite{su_effects_2017,hutchins-delgado_characterization_2022,valentini_parity-conserving_2024}. Further studies of this material indicaties that the MBE-grown material is interface limited\cite{dong_growth_2026}. The mean free path ($l_{\mathrm{mfp}}$) of the Ge QW at different carrier densities is also shown in Fig. \ref{fig:1}c). The maximum estimated $l_{\mathrm{mfp}}$ from the Ge QW is 482 nm. This length places an upper bound on the length of the tunnel region and confirms that it is feasible to fabricate JJs that support ballistic supercurrent to travel across the device.

\begin{figure*}[ht]
\centering
\includegraphics[width=\textwidth]{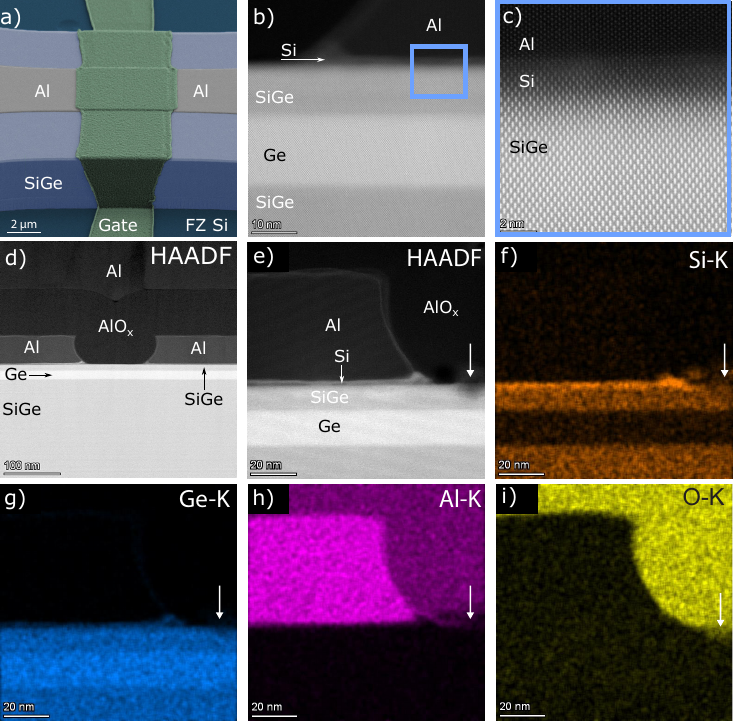}
\caption{\label{fig:2} a) Representative false colored SEM micrograph of a JJ, the gate electrode can be observed climbing the mesa to the JJ gap. b) HAADF STEM of the JJ device. c) \& d) atomic resolution HAADF STEM images of the SiGe/Ge and Si/Al interfaces. e)-i) EDS mapping of the device. The arrow indicates a damaged region of the device from prior EDS mapping.}
\end{figure*}

The fabrication process for a gated JJ begins by defining the junction. The junction is defined by using a BCl\textsubscript{3} (38~sccm) - Cl\textsubscript{2} (2~sccm) inductively coupled plasma (ICP) reactive ion etch (RIE) to remove most of the Al within the junction. The etch is performed at 7~mTorr, 20~\unit{\degreeCelsius}, 100-W of RF power, and 750-W of ICP power. To complete the etch, Microposit MF-319 photo-developer, which contains 2.2 \% tetramethylammonium hydroxide (TMAH), is used to selectively remove the remaining Al without plasma damage to the semiconductor surface. This approach also limits lateral etching of the Al compared to a strictly wet-etch recipe and helps to improve uniformity along the junction. A $\sim$ 2.6-\unit{\um} tall mesa is defined using ICP-RIE with a gas flow of BCl\textsubscript{3} (15 sccm)-O\textsubscript{2} (2 sccm)- Ar (45 sccm) at 20 \unit{\degreeCelsius} and pressure of 4 mTorr. The plasma power is set to 600 W (ICP) and 120 W (RF). The deep mesa etch produces a tapered sidewall with an angle of 18\unit{\degree} from normal. The side wall taper is sufficient to allow metal leads, such as gates or interconnects, to continuously climb to the top of the mesa. After the mesa etch, an ALD gate dielectric was deposited. Before deposition, the sample is cleaned using an \textit{in-situ} O\textsubscript{2} plasma, after which a 50 nm thick ALD AlO\textsubscript{x} layer is deposited at 200 \unit{\degreeCelsius}. Planetary deposition of an Al gate electrode is then used to ensure that it is continuous from the bottom of the mesa, where the bonds pads are located, to the top of the mesa, where the device is located. A scanning electron micrograph (SEM) image of a fabricated JJ is shown in Fig. \ref{fig:2}a), where the continuity of the gate electrode can be observed. This process enables continuous metal leads from the low-loss substrate to traverse the tall mesa to the junction, thereby providing a simple path for integration of our JJs with other low-loss superconducting circuitry on the Si substrate.

To gain insight into the nanometer-scale details of the fabricated JJ, high angle annular dark field (HAADF) scanning transmission electron microscopy (STEM) was performed on fabricated devices. Fig \ref{fig:2}b) \& c) shows HAADF STEM images of the Si/Al and SiGe/Ge interfaces. The lower Ge interface is 3.7~nm-wide, while the top Ge interface is 3.1~nm-wide resulting from growth kinetics of Ge incorporation\cite{zalm_ge_1989, croke_evidence_1990}. The region shown in Fig \ref{fig:2}c) is indicated by the blue box in Fig \ref{fig:2}b) which shows the abrupt interface between the polycrystalline Al film and Si capping layer. Fig. \ref{fig:2}d) shows a lower magnification image of the JJ. No residue or damage is observed in the junction after device processing. The two-step junction etch of the Al produces a tapered sidewall profile of the Al, with the bottom of the junction having a gap of 100 nm, and the top of the junction having a gap of 145 nm. From the higher magnified image of the junction in Fig. \ref{fig:2}e), the Si cap is absent in the junction after device fabrication and may have been removed during the wet etch of the junction gap. Some accumulation of Si and Ge is seen at the bottom edge of the Al electrode for this device (indicated by arrows in Fig. \ref{fig:2}e-i). Energy dispersive spectroscopy (EDS) mapping of the junction was used to investigate possible chemical intermixing. Figure \ref{fig:2}f-i) shows that no chemical intermixing is observed between the Al/Si and SiGe/Ge interface. 

\begin{figure*}
\centering
\includegraphics[width=\textwidth]{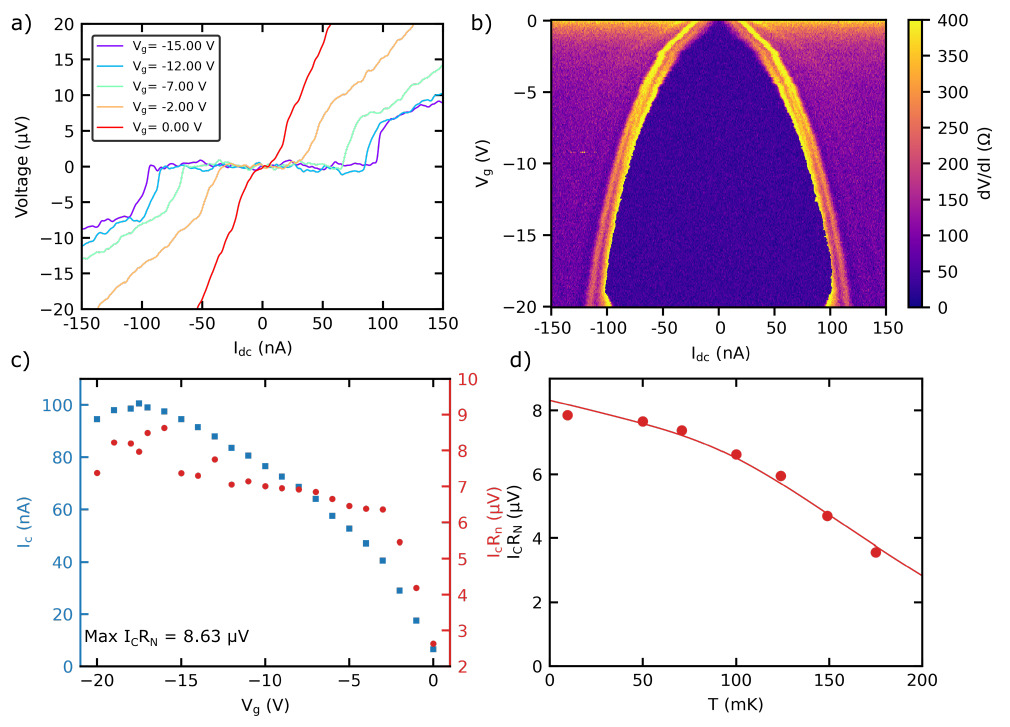}
\caption{\label{fig:3} a) Current-voltage curves measured at 10~mK from a Josephson junction demonstrating gate-tunable superconductivity. b) Color plot of differential resistance of the junction at different DC bias currents and gate voltage. Without an applied gate voltage, there is no apparent supercurrent, as the gate voltage becomes more negative, a supercurrent appears and increases to a maximum of $\pm$103~nA. c) A plot of the critical current normal resistance product. d) Temperature dependence of the junction’s critical current (marks) and fitting of the temperature dependence to the generalized Kulik-Omelyanchuk model (line). V\textsubscript{g} = -17.5 V.}
\end{figure*}

Direct current (DC) transport measurements of the JJs were performed using a dilution refrigerator with cryogenic low-pass filters at a temperature near 10~mK. The differential resistance was measured with a lock-in amplifier using a low-frequency excitation current of 1~nA. The DC current vs voltage (I-V) behavior as a function of gate voltage is shown in Fig. \ref{fig:3}a). The I-V curves demonstrate gate-tunable superconductivity in the JJ, with the supercurrent increasing and the normal state resistance decreasing with more negative gate voltages, which is expected in systems with holes as the majority carriers. 

Fig. \ref{fig:3}b) shows the dependence of the differential resistance (dV/dI) on the current bias and the gate voltage over a wider range of gate voltages from 0 to -20 V. At $V_g$ = 0 V, there is little or no supercurrent, and as $V_g$ decreases, the carrier density in the quantum well increases and a superconducting region begins to open up (differential resistance = 0 $\mathrm{\Omega}$). For this device, the critical current increases to a maximum of 103 nA at $V_g$ = -17.5~V. After this maximum, the critical current begins to decrease with $V_g$. The decrease in critical current is likely due to the mobility of the QW decreasing at very large gate voltages due to the quantum-confined Stark effect\cite{jana_stark-effect_2011} or intersubband scattering \cite{rossner_2-d_2006}.

The voltage-dependence of the carrier transport across the JJ can also be expressed as the $I_CR_N$: a figure of merit commonly used to evaluate the transparency of weak link JJ devices \cite{mayer_superconducting_2019}. Fig. \ref{fig:3}c) displays the dependence of $I_C$ and $I_CR_N$ on Vg for this JJ. The maximum $I_CR_N$ of 8.63 \unit{\uV} coincides near the maximum of the critical current. This $I_CR_N$ value is low compared to the ideal $I_CR_N$ of $\pi\Delta$, and is comparable to the $I_CR_N$ products obtained from other super-semi JJs fabricated with \textit{ex-situ} Al on deep Ge QWs\cite{vigneau_germanium_2019,hendrickx_ballistic_2019}. These observed values are indicative of poor interface transparency of the superconducting contacts to the Ge QW. The interface transparency can be further quantified by studying the temperature dependence of the $I_CR_N$ product as shown in Fig. \ref{fig:3}d). The superconductor-semiconductor interface transparency can be determined with the $I_CR_N$ product temperature dependence and the generalized Kulik-Omelyanchuk model\cite{haberkorn_theoretical_1978}: 

\begin{eqnarray*}
\begin{split}
    I_{C}R_{N} &= \frac{\alpha \pi \Delta (T)}{2e} \frac{sin(\phi)}{ \sqrt{1- \tau sin^2(\frac{\phi}{2})}}\\
    &\times tanh\left( \frac{\Delta(T)}{2 k_B T} \sqrt{1- \tau sin^2 (\frac{\phi}{2})} \right)
\end{split}
\end{eqnarray*}

Where $\phi$ is the phase across the junction, $\alpha$ is a scaling parameter between zero and unity, $\tau$ is the transparency of the junction, \textit{e} is the electron charge. The temperature dependent superconducting gap is calculated using, $\Delta(T)=\Delta_{0}\sqrt{1-{T}/{T_c}}$, the zero-temperature superconducting gap, $\Delta_{0}=1.764k_BT_c$, and the superconducting critical temperature of the metal, $T_c$. Model fitting is completed with the phase maximization of $I_C$ procedure\cite{lee_ultimately_2015}. The extracted interface transparency of approximately 4--9 \% is typical from these devices. The poor transparency is consistent with the low $I_CR_N$ product. Higher $I_CR_N$ products close to the ideal value of $\pi\Delta$ were reported for Ge-QW JJs with a thinner, 8-\unit{\nm} top spacer (3-\unit{\nm}-thinner than this device) and \textit{ex-situ} deposited Al contacts\cite{valentini_parity-conserving_2024}.

To enable a voltage-tunable transmon qubit, the observed critical current of these devices defines the Josesphson energy, $E_J=I_C\Phi_0/2\pi$, where $\Phi_0=h/2e$ is the superconducting magnetic flux quantum and h is the Planck constant. The transmon frequency $f_q\approx(\sqrt{8E_JE_C}-E_C)/h$, where $E_C=e^2/(2C)$ is the singe-electron charging energy associated with the total capacitance $C$ shunting the junction\cite{krantz_quantum_2019}. Given typical design values of transmon qubits with $E_C=200$~ MHz, the measured values of the critical current, $25 < I_C < 100$~nA would enable qubit tuning between 4.25~GHz and 8.7~GHz with $E_J/E_C$ ranging from 62 to 348. These values are within the transmon regime\cite{koch_charge-insensitive_2007}. The upper limit of tuning would most likely be a design consideration regarding the minimum design anharmonicity that was desired instead of a physical limit from the maximum $I_C$ that the JJ might express.

 In conclusion, a deep mesa etch process on MBE grown shallow Ge QW heterostructures using float-zone Si substrates and an \textit{in-situ} deposited Al superconductor is used to make lateral superconductor-semicondcutor Josephson junctions. The clean and abrupt superconductor-semiconductor interface enables proximitization without annealing, prevents uncertainty in the junction length from diffusion, and results in a Josephson junction with a strong gap. The shallow QWs exhibit a mean free path exceeding 400 nm, gate tunability of the critical current exceeding 100 nA, an $I_CR_N$ product of 8.63~\unit{\uV}, and calculated transparency of approximately 4--9 \%. Further improvements in the layer structure are expected to improve the transparency. The fabricated devices demonstrate the essential elements of a design for monolithic integration of voltage-tunable JJs with low-loss superconducting circuits.

\section*{Acknowledgments}
B. B., S. D., and H. O. H. C. acknowledge support from ARO award no. W911NF2220036. This research was funded in part under Air Force Contract No. FA8702-15-D-0001. The views and conclusions contained herein are those of the authors and should not be interpreted as necessarily representing the official policies or endorsements, either expressed or implied, of the U.S. Government.

\section*{Author Declarations}
\subsection*{Conflict of Interest}
The authors have no conflicts to disclose.

\section*{Data Availability}
The data that support the findings of this study are available on request from the corresponding author.

\section*{References}

\bibliography{references}

\end{document}